\documentclass{elsart}

 \usepackage{epsfig}
\def\simlt{\ \raise -2.truept\hbox{\rlap{\hbox{$\sim$}}\raise5.truept   %
\hbox{$<$}\ }}                                                          %
\def\simgt{\ \raise -2.truept\hbox{\rlap{\hbox{$\sim$}}\raise5.truept   %
\hbox{$>$}\ }}                                                          %

\def\be{\begin{equation}}
\def\ee{\end{equation}}
\def\newline{\hfil\break}

\def\szkin{ SZ$_{kin}$ }
\def\szth{ SZ$_{th}$ }

\usepackage{amssymb}
\begin{document}

\begin{frontmatter}



\title{Beyond the Standard Lore of the SZ effect}


\author{S. Colafrancesco}

\address{INAF - Osservatorio Astronomico di Roma \\
 Via Frascati 33, I-00040 Monteporzio, Italy\\
 Email: Sergio.Colafrancesco@mporzio.astro.it}

\begin{abstract}
Multi-frequency (X-ray, optical and radio) observations of galaxy clusters indicate that
the atmospheres of these cosmic structures consist of a complex structure of thermal (hot
and warm) and non-thermal (with different origin and spectra) distribution of electrons
(and protons) which is, therefore, far from its modelling as a single, thermal electronic
gas. This evidence requires to go beyond the simple, standard lore of the SZ effect. This
task is challenging for both the theoretical aspects of their modelling and for the
experimental goals to be achieved, but it will return a large amount of physical
information by using the SZ effect as a unique tool for astro-particle and cosmology.
\end{abstract}

\end{frontmatter}


\section{The SZ effect: the standard lore}
 \label{sec.sz_standard}

The Sunyaev-Zel'dovich effect (hereafter SZE, Zel'dovich \& Sunyaev 1969, Sunyaev \&
Zel'dovich 1972, 1980) produces distortions of the CMB spectrum by means of the Compton
scattering of CMB photons off the energetic electrons which are present in the atmosphere
of cosmic structures, like clusters of galaxies and galaxies (see Rephaeli 1995,
Birkinshaw 1999, Colafrancesco 2004a for reviews).
Such a scattering is proportional to the energy density of the electron population and
produces a systematic shift of the CMB photons from the Rayleigh-Jeans (RJ) to the Wien
side of the spectrum. In this respect, it is a powerful probe of the physical conditions
of electronic plasmas in astrophysical and cosmological context.\\
The standard description of the non-coherent Compton scattering of an isotropic Planckian
radiation field by a non-relativistic Maxwellian electron population -- like the one
constituting the hot (with temperature $T_e \sim 10^7-10^8$ K), optically thin (with
density $n_e \sim 10^{-3} - 10^{-2}$ cm$^{-3}$, and size $R\sim$ Mpc) intracluster (IC)
medium -- can be obtained by means of the solution of the Kompaneets (1957) equation.
As such, the origin of the SZE can be considered as a fall-out effect of the cold war: in
fact, the Compton scattering Fokker-Planck equation for a population of scattered photons
was first derived by A.S. Kompaneets in the early 1950 and then remained classified due
to nuclear bomb research until 1956 (see, e.g., Goncharov 1996); it was finally published
in 1957 (Kompaneets 1957).
In 1969, Ya.B. Zel'dovich and R. Sunyaev (1969) derived the SZ effect applying the
Kompaneets equation to the case of the Compton scattering of CMB photons off the thermal
population of electrons confined in the potential wells of galaxy clusters, i.e. the
intra-cluster medium (ICM).\\
The change in the spectral intensity of the CMB seen in the direction of a galaxy cluster
as due to the scattering of CMB photons by a {\it thermal} electron distribution can be
written as
 \be
\Delta I_{th} = 2{(k_BT_0)^3 \over (hc)^2} y_{th} g(x) ~,
 \label{eq.deltai_th}
 \ee
where $x = h \nu / k_BT_0$ is the a-dimensional frequency and the spectral features of
the effect are contained in the function $g(x)$ which reads as
 \be
g_{non-rel}(x) = {x^4 e^x \over (e^x -1)^2} \bigg[x {e^x +1 \over e^x -1} -4\bigg]
 \label{eq.gdx_th}
 \ee
in the non-relativistic limit.
\begin{figure}[!ht]
\begin{center}
 \epsfig{file=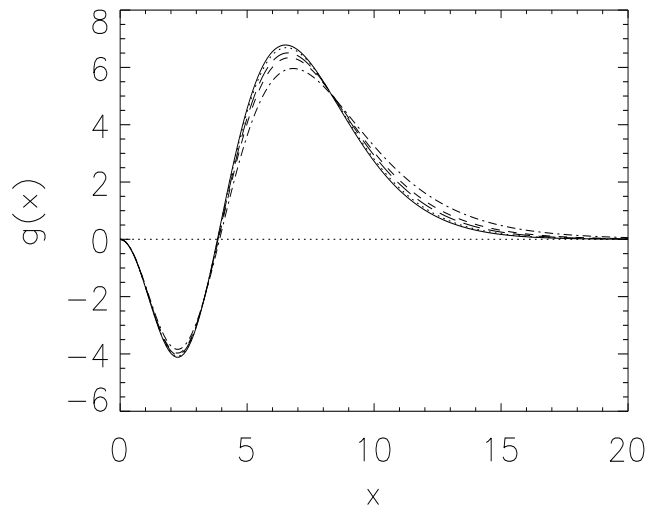,width=10.cm,height=8.cm,angle=0.}
  \caption{\footnotesize{The function $g(x)$ in eq.(2) (solid line) is compared with the function
  $ \tilde{g}(x)$ for thermal electron populations with $k_B T_e=10$ (dot-dashed),
  $5$ (dashes), $3$ (long dashes) and  $1$ (dotted) keV, respectively.
 }}
  \label{fig.gtilde_th}
\end{center}
\end{figure}
This function (see Fig.\ref{fig.gtilde_th}) is zero at the frequency $x_0=3.83$ (or $\nu
= 217$ GHz for a value of the CMB temperature $T_0 = 2.726$ K), negative at $x < x_0$ (in
the RJ side) and positive at $x > x_0$ (in the Wien side). The Comptonization parameter,
$y_{th}$, due to the thermal SZE reads as
 \be
y_{th}= {\sigma_T \over m_e c^2} \int d\ell ~n_e k_BT_e ~,
\label{y_th}
 \ee
where $n_e$ and $T_e$ are the electron density and temperature of the IC gas,
respectively, $\sigma_T$ is the Thomson cross section, valid in the limit $T_e \gg T_0$,
$k_B$ is the Boltzmann constant and $m_ec^2$ is the rest mass energy of the electron. The
Comptonization parameter $y_{th}$ is proportional to the integral along the line of sight
$\ell$ of the kinetic energy density of the IC gas, $\epsilon_{th} \approx n_e k_B T_e$,
or equivalently the kinetic pressure
\footnote{Pressure in a fluid may be considered to be a measure of energy per unit volume
or energy density. According the kinetic theory of ideal gases - see Halliday, Resnick \&
Walker 2000 - the gas pressure can be defined as the average momentum transfer per unit
area per unit time due to collisions between a confined gas and its boundary. Using
Newton's second law, this pressure can be shown to be given by one third of the average
kinetic energy of particles in the gas $\langle \epsilon_{th} \rangle = 3 k_BT_e n_e$.}
that we define here as $P_{th} = n_e k_B T_e$.
Hence, eq.(\ref{y_th}) can be written as
 \be
y_{th} = {\sigma_T \over m_e c^2} \int d\ell ~P_{th} ~,
 \ee
where the relevant dependence from the total kinetic pressure, $P_{th}$, of the electrons
along the line of sight $\ell$ appears.
Since the previous description of the thermal SZE is obtained under the Kompaneets
approximation and in the single scattering regime of the photon frequency redistribution
function, it only provides an approximation of the SZE for low electronic temperatures
($k_B T_e \simlt 3$ keV) and low values of the optical depth
 \be
\tau = \sigma_T \int d\ell n_e \, ,
 \label{eq.tau}
 \ee
which is usually $ \simlt 10^{-3}$ in galaxy clusters.\\
The existence of many high-temperature X-ray clusters (see Arnaud 2005 for a recent
review) with $k_B T_e$ up to $\sim 17$ keV (see, e.g., Tucker et al. 1998, Liang et al.
2002) which correspond to $k_BT_e /m_ec^2 \approx 3.3 \cdot 10^{-2}$, requires to take
into account the appropriate relativistic corrections in the calculation of their thermal
SZE (see Rephaeli 1995, Birkinshaw 1999 and references therein).\\
Analytical and numerical expressions for the SZE in the relativistic case have been
considered by various authors using both analytical and Monte Carlo techniques (see,
e.g., Stebbins 1997; Itoh et al. 1998; Challinor \& Lasenby 1998; see also Birkinshaw
1999 and references therein). Some of these calculations (out of the Monte Carlo
simulations) are still approximate since they have been carried out in the limits of {\it
i)} single scattering of CMB photons against the IC gas electrons; {\it ii)} diffusion
limit in which the use of the Kompaneets equation is justified.
The results of Itoh et al. (1998) based on a generalised Kompaneets equation and by
direct integration of the Boltzmann collision term, can be regarded as exact in the
framework of the single scattering approximation. Analytical fitting formulae of such
derivation are available (Nozawa et al. 2000; Itoh et al. 2002) and offer a detailed
description of the thermal SZE for $k_B T_e \simlt 15 $ keV, while Monte Carlo
simulations describe more correctly the thermal SZE even for $k_B T_e \simgt 20$ keV
(see, e.g., Challinor and Lasenby 1998).  Numerical solutions for the thermal SZE which
are valid for generic values of $\tau$ and $T_e$ have been given by Dolgov et al. (2001)
based on an analytical reduction of the collision integral which is contained in the
Boltzmann-like collision equation.
Sazonov and Sunyaev (2000) presented a derivation of the monochromatic redistribution
function in the mildly relativistic limit which considers also quantum effects and the
use of the Klein-Nishina cross-section which reproduces, in the limit $h \nu \ll k_BT_e$,
the results of Fargion et al. (1997). However, they still consider only the single
Compton scattering limit and the relativistic corrections up to some intermediate order
due to low-energy photons and relativistic electrons.
Itoh et al. (2001) also presented a calculation of the thermal SZE which considers the
contribution from multiple scattering in the relativistic limit, and Colafrancesco et al.
(2003) derived a general form of the SZE valid in the full relativistic regime, for
generic values of $\tau$ and $T_e$ and for multiple scattering regimes.\\
Another general assumption which is made in the calculation of the SZE is the use of a
single population of thermal electrons (i.e., the hot ICM). This assumption is based on
the evidence that the ICM is mainly constituted by thermal electrons (and protons) which
are responsible for the thermal bremsstrahlung X-ray emission observed in clusters (see
Sarazin 1988 for a review). A further assumption is to use the electronic temperature
$T_e$ as a measure of the average energy per particle, a condition which is not ensured
in plasma undergoing fast, non-equilibrium processes which may yield $T_e \neq T_p$.\\
The study of the thermal SZE, caused by the random scatterings of the thermal
(isotropically distributed) electrons, is complemented in the standard description by a
kinematic (Doppler) SZE, \szkin, which appears when the cluster has a whole has a finite
(peculiar) velocity $V_r$ in the CMB frame.
The expressions for the intensity and temperature changes due to the \szkin can be
obtained, assuming that the \szth and \szkin are separable, by a simple relativistic
transformation (see, e.g., Rephaeli 1995 for a review) which yields
 \be
 \Delta I_{kin}(x) = - 2 {k_{\rm B} T_{CMB})^3 \over (h c)^2} {V_r \over c} \tau
 {x^4 e^x \over (e^x -1)^2} \, ,
 \label{eq.szkin}
 \ee
where the line-of-sight peculiar velocity $V_r$ is positive (negative) for a receding
(approaching) cluster.
At variance with the temperature change due to the thermal SZE (see eqs.
\ref{eq.deltai_th} and \ref{eq.gdx_th}), the \szkin temperature change $\Delta T_{kin}= -
T_0 (V_r/c) \tau$ is independent of frequency.

\newline
{\bf The standard lore of the SZE: simple physics and cosmology}\\
Simple astrophysical and cosmological results can be obtained from the study of the
standard (both thermal and kinematic) SZE (see Birkinshaw 2003).\\
As for cluster physics, the SZE can be used to study:
\begin{itemize}
\item the integrated SZE, which provides information on the total thermal energy content
and the total electron content;
\item the spatial SZ structures: these are not as sensitive as the available X-ray data, and also
need for IC gas temperature estimates;
\item the mass structures and their relationship to gravitational lensing derived
structures;
\item the radial peculiar velocity of galaxy clusters via the \szkin .
\end{itemize}
\newline
As for cosmology, the SZE can provide information on:
\begin{itemize}
\item cosmological parameters, like the cluster-based Hubble diagram or the distribution of the
cluster counts vs. redshift, which can be used to probe $\Omega_m$, with a minor
sensitivity to $\Omega_{\Lambda}$ (see Carlstrom et al. 2002);
\item cluster evolution physics, i.e. the evolution of cluster atmospheres, the evolution of
their radial velocity distributions and the evolution of their baryon fraction;
\item the evolution of $T_{CMB}$ with redshift (see, e.g., Battistelli et al. 2002).
\end{itemize}

We must keep in mind, however, that two basic working approximations are assumed in such
use of the standard SZE:
i) the diffusion limit (i.e., the single scattering approximation valid for $\tau \ll
1$);
ii) a single electron population (i.e., the population of thermal hot electrons which is
confined in the cluster atmosphere and emits in the X-rays by bremsstrahlung.\\
There are several pieces of evidence and physical arguments that require to go beyond
this standard lore of the SZE.

\section{The SZE: more than basics.}
 \label{sec.beyond}

The electronic distribution of the atmospheres of galaxy clusters is neither simple nor
unique.
\begin{figure}[!ht]
\begin{center}
\hbox{
 \epsfig{file=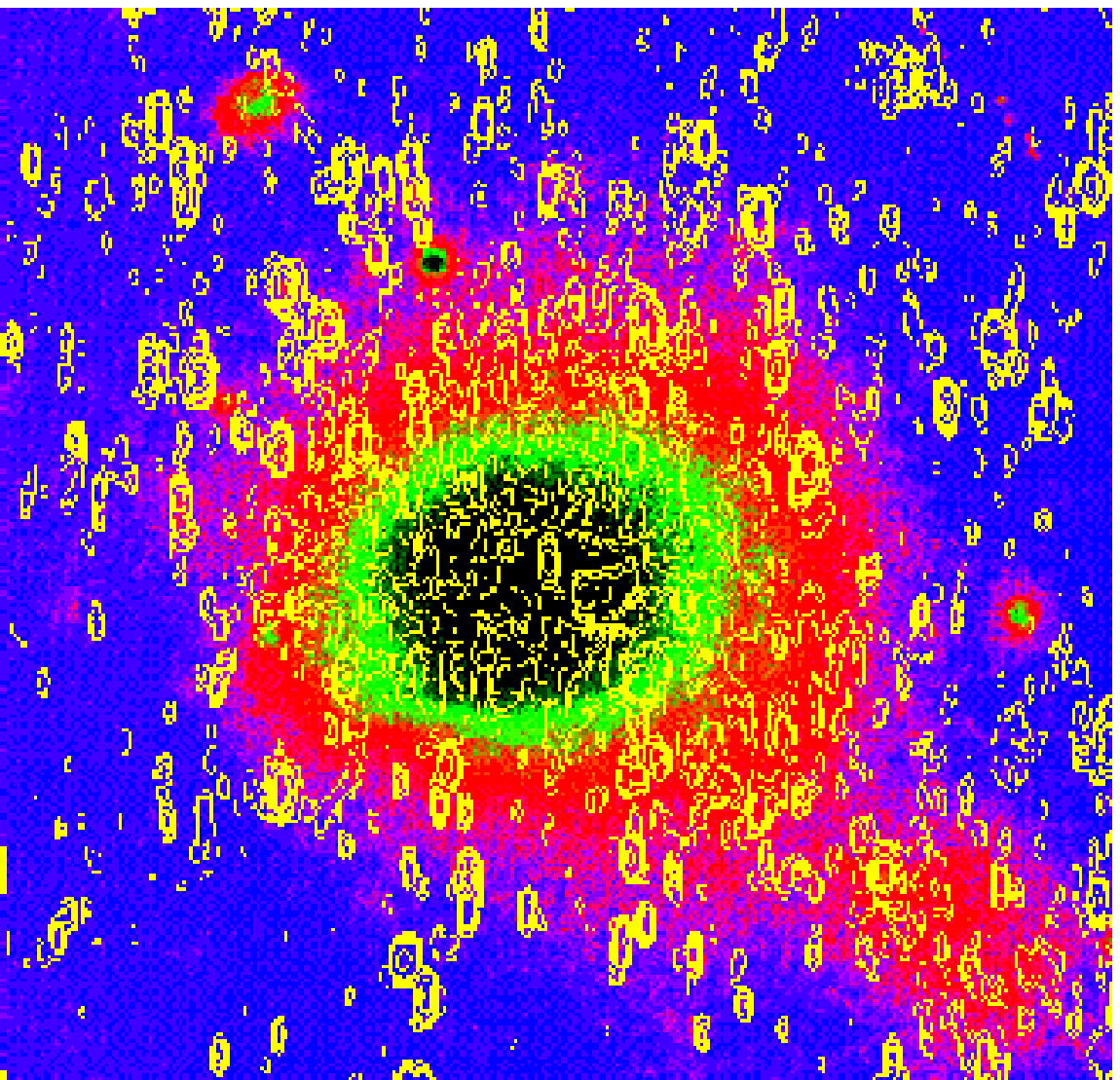,width=4.5cm,angle=0.}
 \epsfig{file=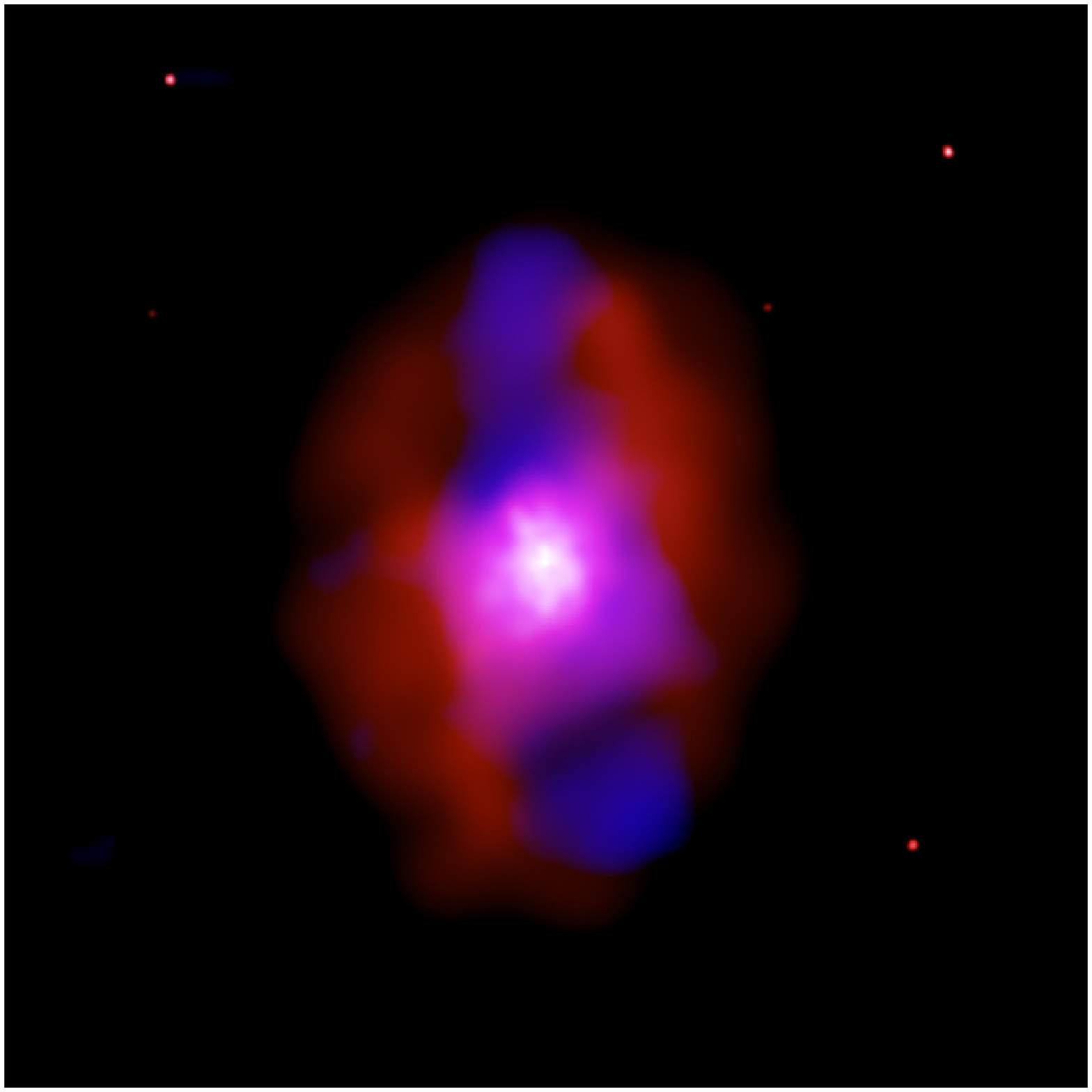,width=4.5cm,angle=0.}
 \epsfig{file=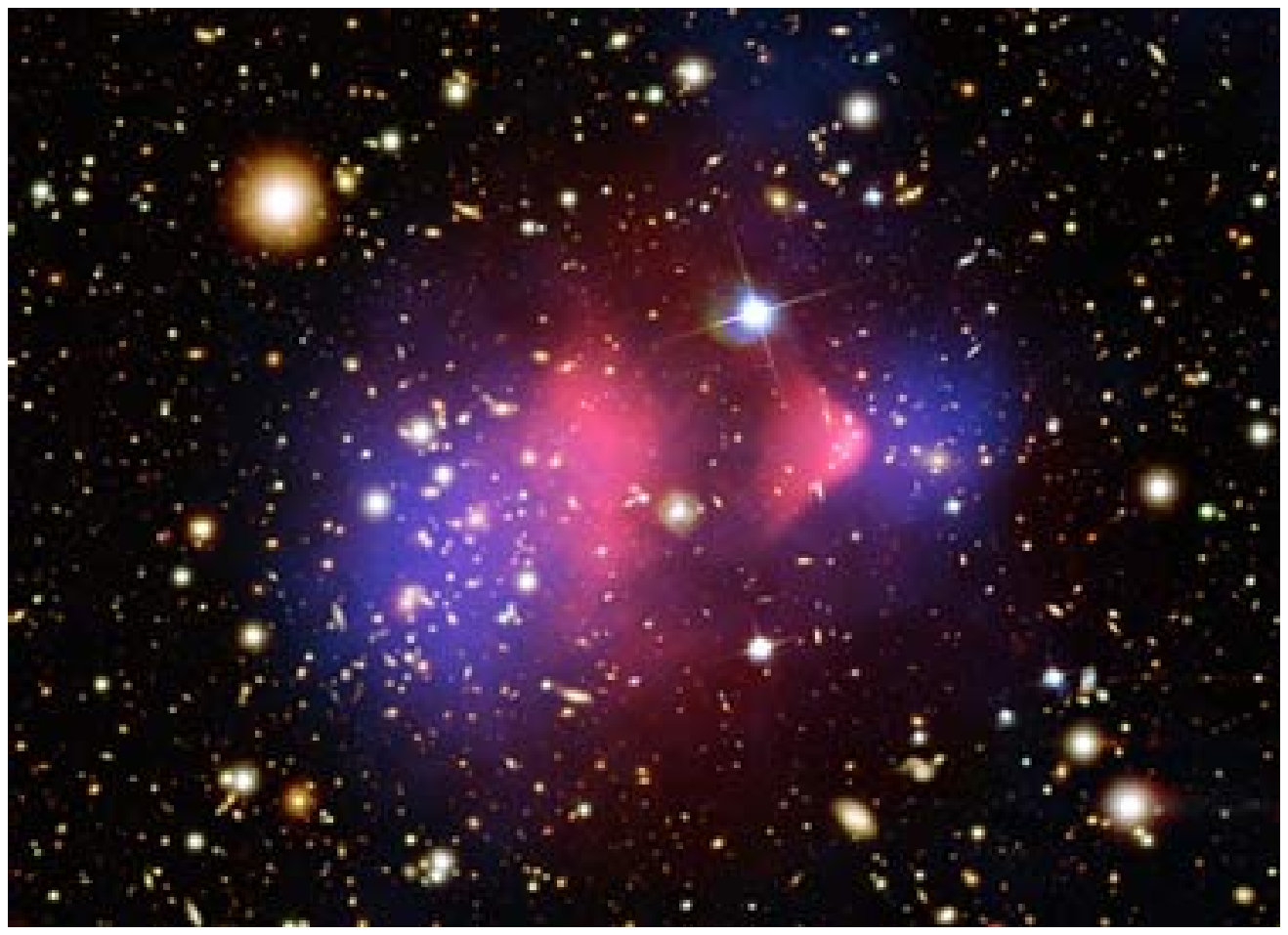,width=4.5cm,angle=0.}
}
  \caption{\footnotesize{We show the images of three clusters which have atmospheres
  with complex electron distributions: Coma with its co-spatial X-ray (colours) and
  radio-halo (contours) emission (left, Feretti 2003), MS0735+7421 with two large cavities
  filled with relativistic plasma (blue) embedded in its thermal (red) IC gas
  (center, McNamara et al. 2005; http://chandra.harvard.edu/photo/2005/ms0735/),
  1ES0657-556 with two DM clumps (blue) offset with
  respect to the complex ICM X-ray (pink) emission (right, Clowe et al. 2006;
  http://hubblesite.org/newscenter/newsdesk/archive/releases/2006/39/image/a).
 }}
  \label{fig.clusters}
\end{center}
\end{figure}
There are, in fact, three matter components in clusters that can provide different
sources of electrons: baryons, Dark Matter, relativistic plasmas.\\
While the SZE from the baryonic, hot plasma component has been described in the previous
Sect.\ref{sec.sz_standard}, we will provide here the basic information on other
non-thermal electronic components residing in cluster atmospheres.\\
Many galaxy clusters contain -- in addition to the thermal IC gas -- a population of
relativistic electrons which produce a diffuse radio emission (radio halos and/or relics)
via synchrotron radiation in a magnetized ICM (see, e.g., Govoni \& Feretti 2004 for a
review). The electrons which are responsible for the radio halo emission have energies
$E_e \approx 16 {\rm GeV} B_{\mu}^{-1/2} (\nu / GHz)^{1/2} \simgt$ a few GeV to radiate
at frequencies $\nu \simgt 30$ MHz in a $\sim \mu$G magnetic field ($B_{\mu} \sim 1$) in
order to reproduce the main properties of the observed radio halos (see, e.g., Blasi \&
Colafrancesco 1999; Colafrancesco \& Mele 2001, Colafrancesco et al. 2006a and references
therein). The origin of such relativistic electrons is not certain and models of
bottom-up production (i.e., re-accelerated by IC turbulence, see e.g. Brunetti 2003 for a
review) or top-down origin (i.e., secondarily produced by Dark Matter WIMP annihilation,
see e.g. Colafrancesco et al. 2006a,b) can fit the observed radio-halo features.\\
The presence of Extreme UV/soft X-ray excesses  (Lieu et al. 1996, Kaastra et al. 2002;
Bowyer 2000) and of an hard X-ray excess (Fusco-Femiano 2004; Rephaeli 2004; Kaastra et
al. 1999) in a few nearby clusters indicate the presence of an additional population of
secondary relativistic electrons (see Bowyer et al. 2004, Marchegiani, Perola \&
Colafrancesco 2006) or a combination of warm (reproducing the EUV excess, Lieu et al.
2000) and (quasi-)thermal (reproducing the hard X-ray excess by bremsstrahlung, see Wolfe
\& Melia 2006, Dogiel et al. 2006) populations of distinct origins.\\
The further evidence for new physical phenomena occurring in the cluster atmospheres --
e.g., non-thermal heating in the cluster cores (see, e.g., Colafrancesco, Dar \& DeRujula
2004 and references therein), AGN and radio-galaxy feedback (Siemiginowska et al. 2005),
intra-cluster cavities (McNamara et al. 2005) and radio bubbles (Birzan et al. 2004)
filled with relativistic non-thermal electrons, multi-scale magnetic fields (see, e.g.,
Govoni \& Feretti 2004) -- imply the presence of additional electronic components with
peculiar spectral and spatial characteristics (see Fig.\ref{fig.clusters}).\\
Finally, there are strong physical arguments indicating that viable Dark Matter candidate
annihilation can produce copious amounts of secondary electrons with a spatial
distribution which, in massive clusters like Coma, is strictly related to that of the
original DM (see Colafrancesco et al. 2006a). It is interesting to note, in this context,
that the spectral distribution of DM-produced secondary electrons carries information on
the mass and the physical composition of the original DM particles.\\
In conclusion, the cluster electronic atmosphere is a complex combination of thermal (hot
and warm) and non-thermal (quasi-thermal due to stochastic acceleration, relativistic due
to DM annihilation and/or to AGN injection) distributions with different energy spectra
and spatial distributions (see, e.g., Colafrancesco 2006b for a review).
Each one of the electron populations which reside in the cluster atmosphere inevitably
produces a distinct SZE with peculiar spectral and spatial features.\\
The description of the non-thermal SZE produced by a single electron population with a
non-thermal spectrum has been attempted by various authors (McKinnon et al. 1991;
Birkinshaw 1999; Ensslin \& Kaiser 2000). Several limits to the non-thermal SZE are
available in the literature (see, e.g., Birkinshaw 1999 for a review) from observations
of galaxy clusters which contain powerful radio halo sources (such as Coma and A2163) or
radio galaxies (such as A426), but only a few detailed analysis of the results (in terms
of putting limits to the non-thermal SZE) have been possible so far (see Colafrancesco et
al. 2003, Colafrancesco 2004a).
The problem of detecting the non-thermal SZE in radio-halo clusters is likely to be
severe because of the associated synchrotron radio emission, which could contaminate at
low radio frequencies the small negative signal produced by the SZE. At higher
frequencies there is in principle more chance to detect the non-thermal SZE, but even
here there are likely to be difficulties in separating the SZE from the flat-spectrum
component of the synchrotron emission (see Birkinshaw 1999). In addition, Colafrancesco
(2004a) noticed that dust obscuration does not allow any detection of the SZ signal from
clusters at frequencies $\simgt 600$ GHz.\\
From the theoretical point of view, preliminary calculations (Birkinshaw 1999, Ensslin \&
Kaiser 2000) of the non-thermal SZE have been carried out in the diffusion approximation
($\tau \ll 1$), in the limit of single scattering and for a single non-thermal population
of electrons.
Matters are significantly more complicated if the full relativistic formalism is used.
However, this is necessary, since many galaxy clusters show extended radio halos and the
electrons which produce the diffuse synchrotron radio emission are certainly highly
relativistic so that the use of the Kompaneets approximation is invalid. Moreover, the
co-spatial presence of thermal and non-thermal electrons renders the single scattering
approximation and the single population approach unreasonable, so that the treatment of
multiple scattering among different electronic populations coexisting in the same cluster
atmosphere is necessary to describe correctly the overall SZE.
In this context, a complete and general derivation of the SZE produced by a general
distribution of non-thermal electrons in the full-relativistic desciption, with multiple
scatterings and also with different families of electrons co-spatially distributed has
been provided by Colafrancesco et al. (2003). We will refer to this approach in our
discussion aimed to probe the origin of every particle family using a single technique:
the SZE.
\begin{figure}[!ht]
\begin{center}
 \epsfig{file=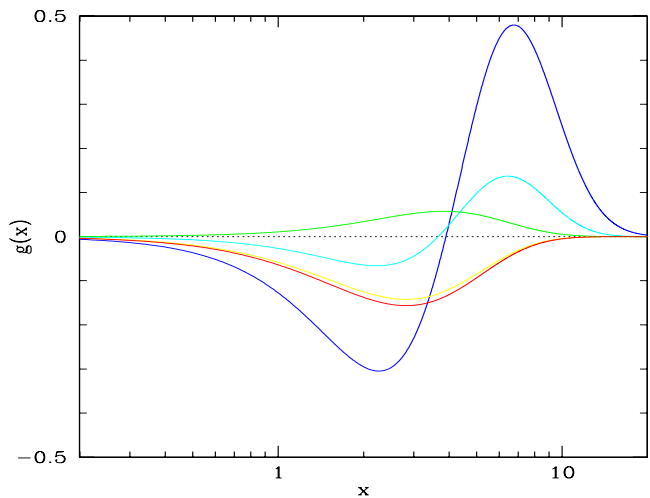,width=9.cm,height=7.cm,angle=0.}
  \caption{\footnotesize{The function $g(x)$ is shown as a function of the
  adimensional frequency $x$ for different electronic populations residing in the cluster
  atmosphere: thermal with $k_B T_e = 8.2$ keV (blue); warm with $k_B T_e = 1$ keV
  (cyan); secondary electrons from DM annihilation with $M_{\chi} = 20$ GeV (red);
  relativistic electrons which fit the Coma radio halo integrated spectrum (yellow).
  Also the kinematic SZE with a negative peculiar velocity (green) is shown for comparison.
  The amplitudes of the various curves have been artificially re-normalized to highlight their
  frequency dependence.
 }}
  \label{fig.gtilde}
\end{center}
\end{figure}

\section{The SZE: a generalized description}

The generalized expression for the SZE  which is valid in the Thomson limit ($\gamma h
\nu \ll m_e c^2$ in the electron rest frame) for a generic electron population, in the
full relativistic treatment and includes also the effects of multiple scatterings and the
combination with other electron populations has been derived by Colafrancesco et al.
(2003) and we will refer to this paper for technical details.
Such derivation is has the advantage to describe both thermal and non-thermal SZE using a
unique formalism in terms of a generalized Compton parameter $y$ and of a spectral
function $\tilde{g}(x)$.\\
According to these results, the spectral distortion observable in the direction of a
galaxy cluster can be written as
 \begin{equation}
\Delta I_{\rm}(x)=2\frac{(k_{\rm B} T_{CMB})^3}{(hc)^2}y_{\rm } ~\tilde{g}(x) ~,
 \label{eq.deltai}
\end{equation}
where the generalized Comptonization parameter $y_{\rm}$ is given by
\begin{equation}
y_{\rm }=\frac{\sigma_T}{m_{\rm e} c^2}\int P_{\rm e} d\ell ~,
 \label{eq.ygen}
\end{equation}
in terms of the pressure $P_{\rm e}$ contributed by the specific electron distribution
within the cluster.
The function $\tilde{g}(x)$ for the considered electron population can be written, in its
most general form, as
\begin{equation}
 \label{gnontermesatta}
 \tilde{g}(x)=\frac{m_{\rm e} c^2}{\langle k_{\rm B}
T_{\rm e} \rangle} \left\{ \frac{1}{\tau} \left[\int_{-\infty}^{+\infty} i_0(xe^{-s})
P(s) ds- i_0(x)\right] \right\} \, ,
\end{equation}
in terms of the photon redistribution function $P(s)$ and of the undistorted CMB spectrum
$i_0(x) = 2 (k_{\rm B} T_{CMB})^3 / (h c)^2 \times x^3/(e^x -1)$. The quantity
\begin{equation}
 \langle k_{\rm B} T_{\rm e} \rangle  \equiv  \frac{\sigma_{\rm T}}{\tau}\int P_e d\ell
= \int_0^\infty dp f_{\rm e}(p) \frac{1}{3} p v(p) m_{\rm e} c
 \label{temp.media}
\end{equation}
(see Colafrancesco et al. 2003) is the analogous of the average temperature for a thermal
electron population (in this case $\langle k_{\rm B} T_{\rm e} \rangle = k_{\rm B} T_{\rm
e}$ obtains, in fact).\\
The photon redistribution function $P(s)= \int dp f_{\rm e}(p) P_{\rm s}(s;p)$ with $s =
\ln(\nu'/\nu)$, in terms of the CMB photon frequency increase factor $\nu' / \nu$,
contains the crucial dependence on the electron momentum distribution $f_{\rm e}(p)$,
where $p$ is normalized to $m_ec$. Here $P_{\rm s}(s;p)$ is the mono-energetic frequency
redistribution function (see Colafrancesco et al. 2003 for details).\\
For a thermal electron population in the non-relativistic limit with momentum
distribution $f_{e,th} \propto p^2 exp(-\eta \sqrt{1+p^2})$ and $\eta= m_e c^2/k_B T_e$,
the pressure writes as $P_{th}=n_e k_B T_e$ and it is easy from eq.(\ref{eq.ygen}) to
re-obtain the Compton parameter
\begin{equation}
  y_{th}=\frac{\sigma_T}{m_e c^2}\int n_e k_B T_e d\ell =\tau \frac{k_B T_e}{m_e
  c^2} ~
\end{equation}
(we consider here, for simplicity, an isothermal cluster with electronic temperature
$T_e$).
The relativistically correct expression of the function $\tilde{g}(x)$ for the same
thermal population of electrons writes, at first order in $\tau$, as:
\begin{equation} \label{gtilde1}
 \tilde{g}(x)=\frac{\Delta i}{y_{th}}=\frac{\tau [j_1-j_0]}{\tau \frac{k_B T_e}{m_e
 c^2}}= \frac{m_e c^2}{k_B T_e}[j_1-j_0] ~,
\end{equation}
where $j_i \equiv J_i \frac{(hc)^2}{2(k_B T_0)^3}$ and the functions $J_i$ are given in
eq.(18) of Colafrancesco et al. (2003). Along the same line, it is  possible to write the
expression of $\tilde{g}(x)$ up to higher orders in $\tau$ (see Colafrancesco et al.
2003).
The expression of $\tilde{g}(x)$ approximated at first order in $\tau$ (as given by
eq.\ref{gtilde1}) is the one to compare directly with the expression of $g(x)$ obtained
from the Kompaneets (1957) equation, since both are evaluated under the assumption of
single electron-photon scattering. Fig.\ref{fig.gtilde_th} shows how the function
$\tilde{g}(x)$ tends to $g(x)$ for lower and lower IC gas temperatures $T_e$. This
confirms that the distorted spectrum obtained from the Kompaneets equation is the
non-relativistic limit of the exact spectrum.
It is worth to notice that while in the non-relativistic case it is possible to separate
the spectral dependence of the effect [which is contained in the function $g(x)$] from
the dependence on the cluster parameters [which are contained in Compton parameter $y$],
this is no longer valid in the relativistic case in which the function $J_1$ depends
itself also on the cluster parameters. Specifically, for a thermal electron distribution,
$J_1$ depends non-linearly from the electron temperature $T_e$ through the frequency
redistribution function $P_1(s)$ (see Colafrancesco et al. 2003 for details). This means
that, even at first order in $\tau$, the spectral shape $\tilde{g}(x)$ of the SZE depends
on the cluster parameters, and mainly from the electron pressure $P_{e}$. This result is
valid for any general electron distribution.

In order to generalize the derivation of the SZE to any arbitrary non-thermal electron
distribution, it is necessary to use the most general expressions of $P_e$ and $f_e(p)$
in eqs.(8) and (9). For example, the general expressions for the pressure of a
relativistic non-thermal distribution of electrons with a power-law spectrum, $n_{e,rel}
= A E^{-\alpha}$, is the following:
\begin{equation}
 \label{press_rel}
 P_{\rm e}=n_{\rm e,rel} \int_0^\infty dp f_e(p) \frac{1}{3} p v(p) m_e c =
   \frac{n_{\rm e,rel} m_e c^2 (\alpha
  -1)}{6[p^{1-\alpha}]_{\rm p_2}^{p_1}}
  \left[B_{\frac{1}{1+p^2}}\left(\frac{\alpha-2}{2},
   \frac{3-\alpha}{2}\right)\right]_{\rm p_2}^{p_1}
\end{equation}
(see, e.g., Ensslin \& Kaiser 2000, Colafrancesco et al. 2003), where $B_x(a,b)=\int_0^x
t^{a-1} (1-t)^{b-1} dt$ is the incomplete Beta function.
For a relativistic population of electrons the relation ${\epsilon}_{e} = 3 \cdot P_{e}$
holds. It is clear from eq.(\ref{press_rel}) that the pressure (and energy density) of
the electron population depends mainly from the value of the minimum momentum $p_1$ for
spectra with slopes $\alpha > 2$.
For an electron population with a double power-law (or more complex) spectrum, analogous
results can be obtained (see Colafrancesco et al. 2003).\\
At first order in $\tau$ the function $\tilde{g}(x)$ of the non-thermal SZE writes as
\begin{equation}
\label{gnonterm1ord}
 \tilde{g}(x)=\frac{\Delta i}{y_{non-th}}=\frac{\tau [j_1-j_0]}{\frac{\sigma_T}{m_e c^2}\int P_e
 d\ell} \equiv \frac{m_e c^2}{\langle k_B T_e \rangle} [j_1-j_0] \, ,
\end{equation}
and the Comptonization parameter $y_{non-th}$ writes, as a function of the quantity
$\langle k_B T_e \rangle$ and of the optical depth $\tau$, as
\begin{equation} \label{y.tmedia}
y_{non-th}=\frac{\sigma_T}{m_e c^2}\int P_{rel} d\ell= \sigma_T \frac{\langle k_B T_e
\rangle}{m_e c^2} \int n_{e,rel} d\ell= \frac{\langle k_B T_e \rangle}{m_e c^2} \tau \, .
\end{equation}
The spectral function of the SZE evaluated for various electron spectra is shown in
Fig.\ref{fig.gtilde}. It is clear that the different nature of the electron distribution
reflects directly in the different spectral shape of the relative SZE. Thus, the SZE is a
powerful probe of the energy spectrum of the various electronic populations residing in
the atmospheres of cosmic structures.
We will describe in the following the relevance of the SZE as a tool to study several
different aspects of the astro-particle physics of cosmic structures in the universe.

\subsection{The SZE and cosmic rays in galaxy clusters}

We know that high-energy and relativistic particles exist in galaxy clusters through
their diffuse radio-halo emission and their hard X-ray emission, but we do not know yet
their origin. There are three viable scenarios for the production of high-energy
particles in the atmospheres of galaxy clusters: i) direct acceleration or stochastic
re-acceleration; ii) injection by AGNs or other compact objects; iii) Dark Matter
annihilation.

\noindent{\bf Testing the acceleration history of cosmic rays}.
One of the most favourite scenarios is that high-energy particles (cosmic rays) are
energized by stochastic acceleration due to e.g., cluster turbulence, merging shocks or
other injection/acceleration mechanisms acting coherently on the large ($\sim$ Mpc) scale
of the cluster atmospheres. Specific models of turbulent re-acceleration producing a
high-energy tail up to relativistic energies (e.g., Brunetti et al. 2003) and stochastic
acceleration of thermal particles producing a quasi-thermal tail (Dogiel et al. 2006)
have been worked out.
\begin{figure}[!ht]
\begin{center}
\hbox{
 \epsfig{file=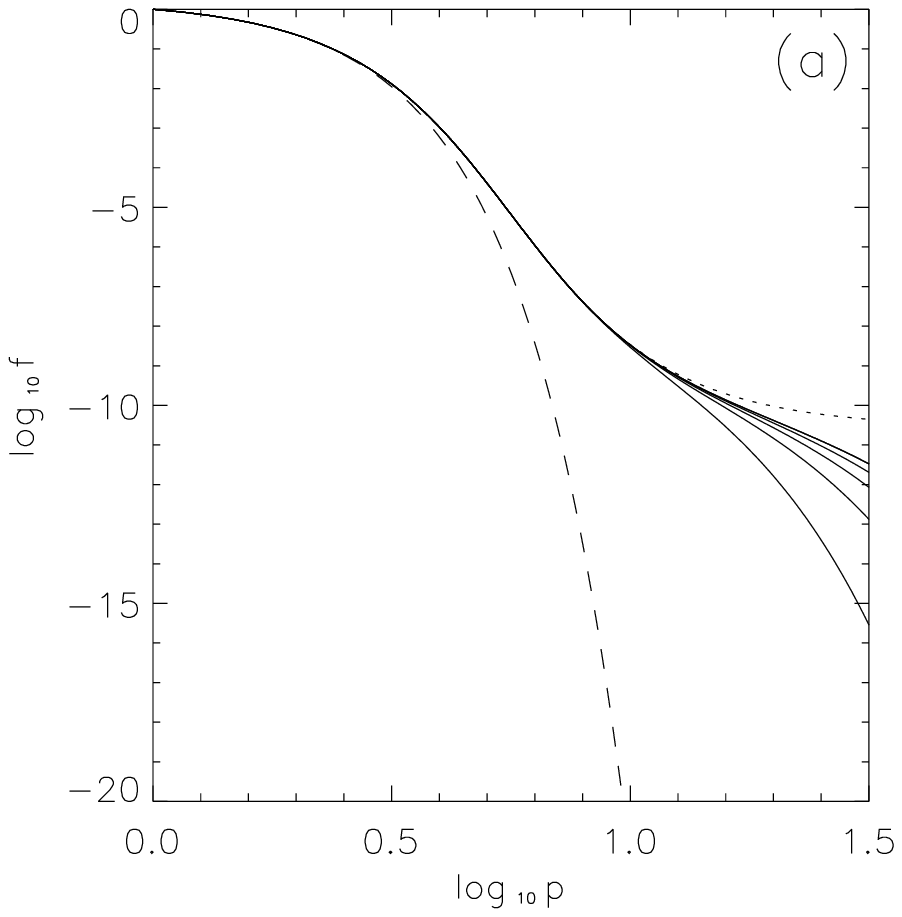,width=6.cm,angle=0.}
 \epsfig{file=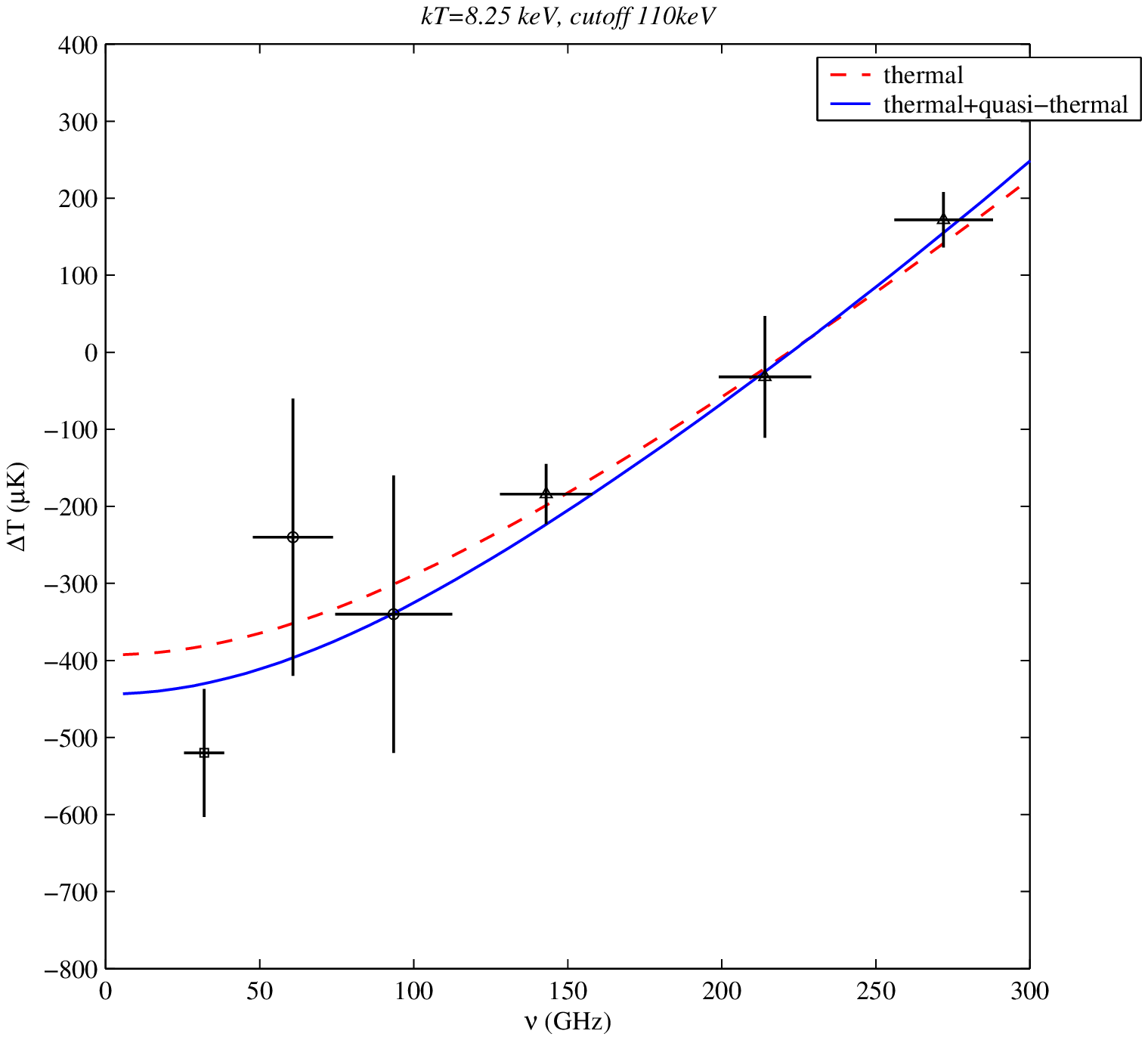,width=6.cm,angle=0.}
  }
  \caption{\footnotesize{The evolution of the electron distribution (left) and the
  CMB temperature change (right) due to the associated SZE for a scenario of
  stochastic acceleration of electrons in the Coma cluster (from Dogiel et al. 2006).
  The dashed curve in the left panel represents the initial Maxwellian distribution while the
  dotted curve is the asymptotic power-law distribution reachable at large times.
 }}
  \label{fig.Coma_stochastic}
\end{center}
\end{figure}
Whatever is the mechanism that accelerate particles, thus producing non-thermal or
quasi-thermal spectral tails, it should be extremely efficient to beat the quite fast
competing thermalization process for such high-energy particles (see, Wolfe \& Melia
2006). As a result, the viable model produce rather different non-thermal or
quasi-thermal electron spectra (see an example in Fig.\ref{fig.Coma_stochastic}) whose
nature can be constrained by studying the associated SZE.\\
If the acceleration is so efficient to produce a high-E tail, the SZE attributed to the
relativistic electrons that produce the cluster radio-halo spectra observed in the 30 MHz
- 5 GHz range is expected to be completely negative at all the frequencies relevant for
the SZ experiments ($\nu \sim 20 - 350$ GHz) and with an amplitude of a few \% of the
thermal SZE in the cluster
(see Colafrancesco et al. 2003, Colafrancesco 2004a, Shimon \& Rephaeli 2003) which
renders its detection quite challenging.\\
On the other hand, the SZE effect produced by quasi-thermal electrons stochastically
accelerated in the cluster environment is expected to produce a detectable CMB
temperature decrement $\Delta T \sim 40-50$ $\mu$K at low frequencies (Dogiel et al.
2006, see also Fig.\ref{fig.Coma_stochastic}) that could prove or set relevant
constraints to this scenario as well as to the efficiency of thermalisation and
acceleration processes.

\noindent{\bf Testing the content and the energetics of cluster cavities}.
Cavities with diameters ranging from a few to a few hundreds of kpc have been observed by
Chandra in the X-ray emission maps of several galaxy clusters and groups (see, e.g.,
Birzan et al. 2004, McNamara et al. 2005). These cavities are supposed to contain
high-energy plasma with a non-thermal spectrum $n_{\rm e,rel} \propto E_e^{-\alpha}$ with
a typical index $\alpha \sim 2.5$. While the properties of these cavities and of the
relativistic plasma they contain is usually studied by combining high-resolution X-ray
and radio maps, we have proposed, as an alternative strategy, to study the consequences
of the Compton scattering between the high-energy electrons filling the cavities and the
CMB photon field (i.e. the SZE)  whose amplitude, spectral and spatial features depend on
the overall pressure and energetics of the relativistic plasma in the cavities (see
Colafrancesco 2005a for the specific case of the cluster MS0735+7421).
\begin{figure}[!ht]
\begin{center}
\hbox{
 \epsfig{file=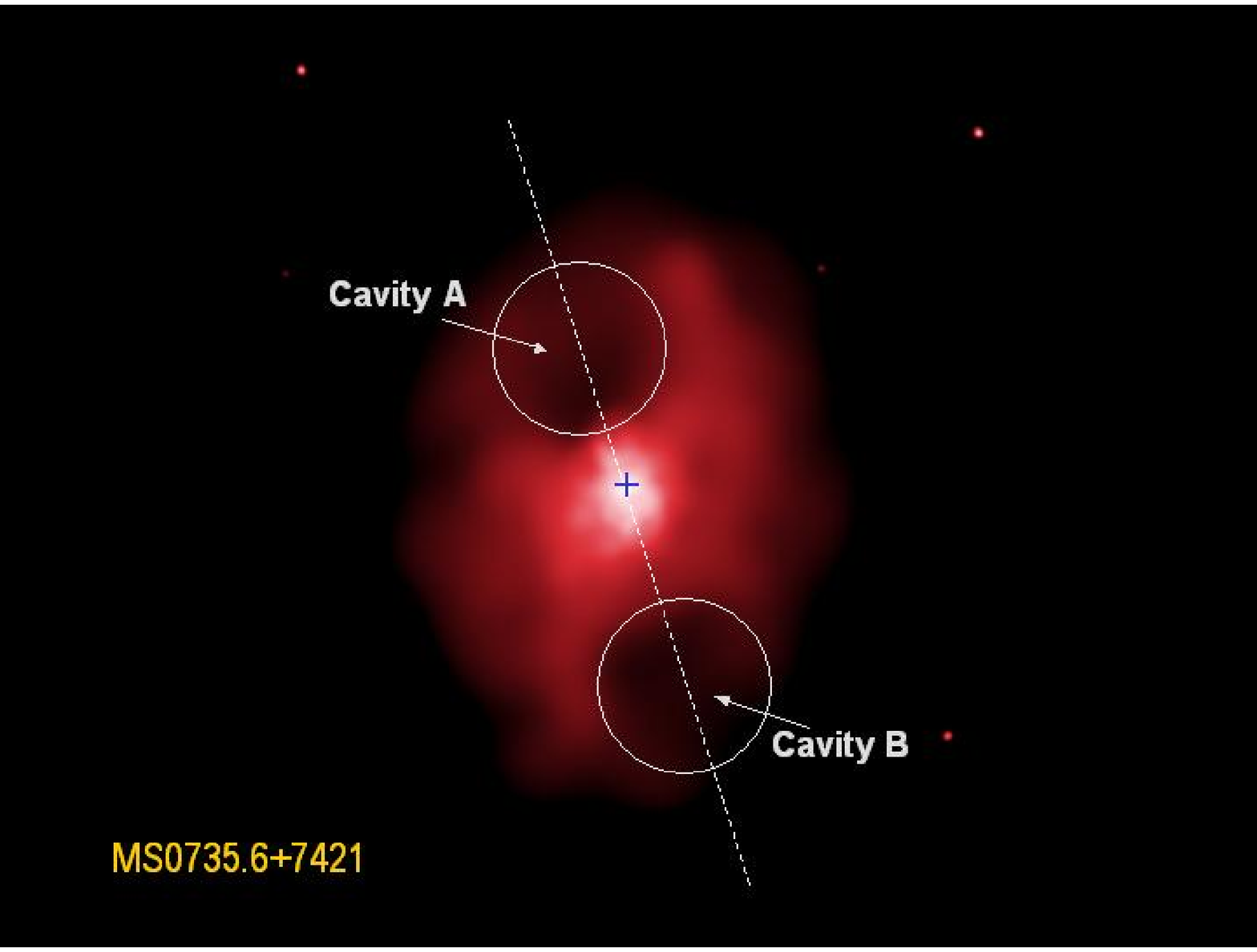,width=6.cm,angle=0.}
 \epsfig{file=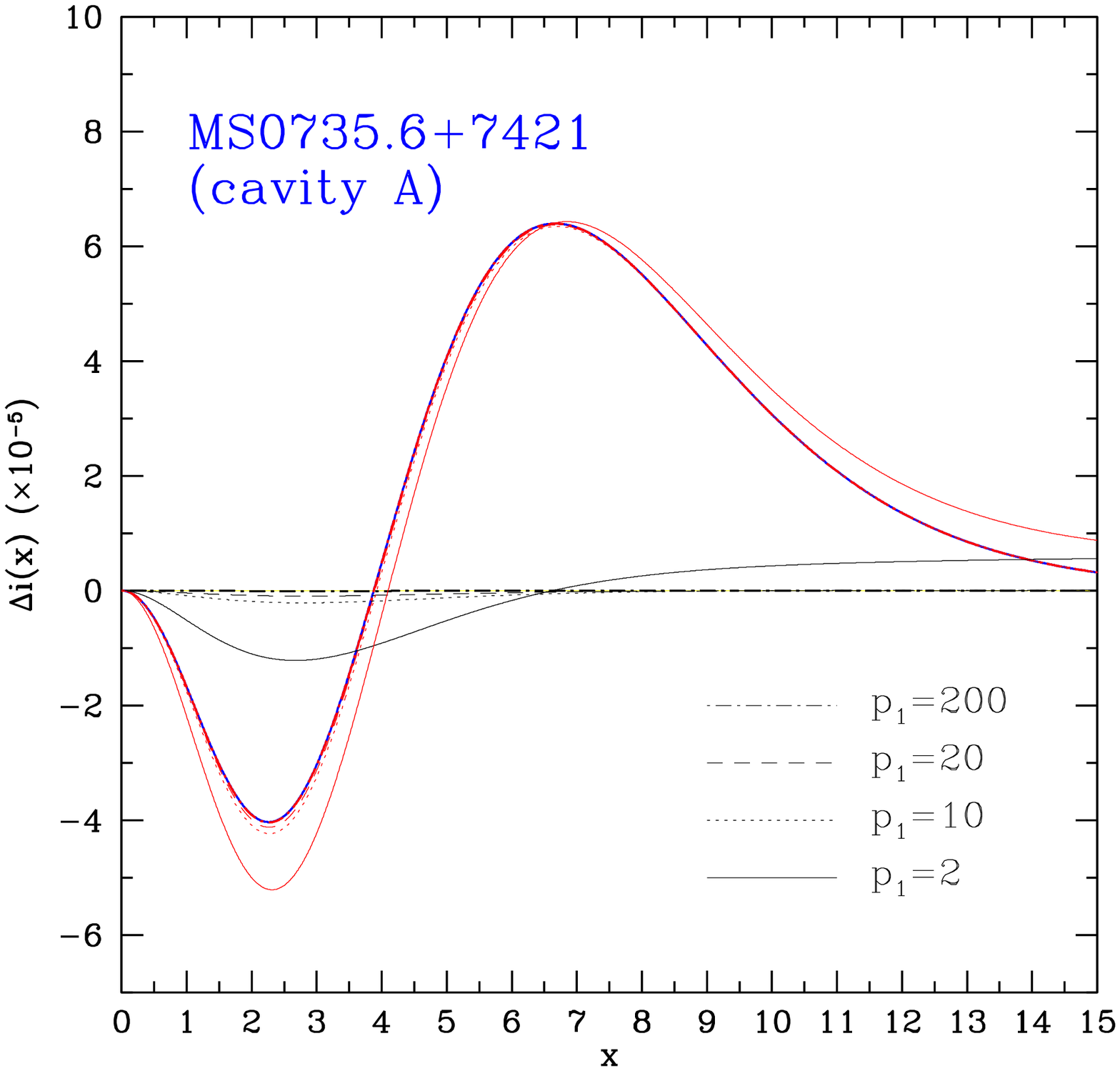,width=6.cm,angle=0.}
  }
  \caption{\footnotesize{Left. The geometry of the cavities in the cluster MS0735.6+7421
  is shown. The two cavities have a radius $\approx 100$ kpc and are located at a distance of
$\approx 125$ kpc and $\approx 170$ kpc from the central radio galaxy (whose position is
indicated by a cross) along the axis represented in the picture.
 Right. The spectrum of the overall SZE from the cluster MS0735.6+7421 has been computed
  at a projected radius of $\approx 125$ kpc from the cluster center where the los passes through the
  center of cavity A. We show the thermal SZE (blue), the non-thermal SZE from the
  cavity (black) and the total SZE (red).
  The plotted curves are for different values of the lowest electron momentum: $p_1= 200$
  (dot-dashes), $p_1=20$ (dashes), $p_1=10$ (dots) and $p_1=2$ (solid). The non-thermal SZE is
  normalized to the cavity pressure $P=6 \times 10^{-11}$ erg cm$^{-3}$ and, in this
  respect, it must be considered as a lower limit of the true SZE coming from the cavity.
 }}
  \label{fig.sz_cavity}
\end{center}
\end{figure}
At frequencies $x \sim 2.5$ there is the maximum amplitude of the non-thermal SZE from
the cavity (for a given electron spectrum and pressure). This produces a bump in the
spatial distribution of the overall SZE at the cavity location with the addition of a
negative SZE signal to the thermal SZE of the cluster (see Colafrancesco 2005a, Pfrommer
et al. 2005).
At $x > x_{\rm 0,th}(P_{\rm th})$ we have the opposite effect but with smaller
amplitudes: a depression in the SZE at the cavity location with the addition of a
negative SZE signal to the positive thermal SZE.
We emphasize that SZ observations at the frequency $x = x_{\rm 0,th}(P_{\rm th})$ (which
is $\approx 3.87$ for a $k_B T=5$ keV cluster) provide a unique tool to probe the overall
energetics, the pressure and the spatial extent of the non-thermal plasma contained in
giant cavities, an observation which is rich in information and complementary to those
obtained by X-ray and radio observations of cluster cavities. At this frequency, in fact,
the overall SZE from the cluster reveals only the Compton scattering of the electrons
residing in the cavities without the presence of the intense thermal SZE observable at
lower and higher frequencies. Hence, the SZE from a cluster containing cavities (like the
case of MS0735.6+7421) shows up uncontaminated at frequencies $\sim 220$ GHz: it is less
extended than the overall cluster SZE because it is only emerging from the cavity regions
and it is also well separable because the cavities are well defined in both X-rays and SZ
images.
We also emphasize, in addition, that the observation of the zero of the non-thermal SZE
in the cavities (which is found at high frequencies, depending on the value of $p_1$ or
equivalently on the value of the pressure in the cavity) provides a definite way to
determine uniquely the total pressure and hence the nature of the electron population
within the cavity, an evidence which adds crucial, complementary information to the X-ray
and radio analysis.
A plausible source of bias to these observations could be provided by a possibly relevant
kinematic SZE due to the cluster peculiar velocity (see Colafrancesco et al. 2003,
Pfrommer et al. 2005 for a discussion).
However, the SZE from the giant cavities in a cluster like MS0735.6+7421 can be
effectively studied at frequency $x = x_{\rm 0,th}(P_{\rm th}) \approx 3.87$ where it is
not affected by the thermal SZE and it is only marginally affected by a possible
kinematic SZE even at a level of a few hundreds km/s (see Colafrancesco 2005a).
Such studies show more generally that the combination of high spatial resolution and
high--sensitivity SZ observations with X-ray and radio data will definitely shed light on
the morphology, on the physical structure, on the dynamics and the origin of these
recently discovered non-thermal features of galaxy clusters. These studies will be also
relevant to determine the impact of specific events of the nature of cavities on the use
of SZ and X-ray clusters as probes for cosmology and for the large scale structure of the
universe.

\noindent{\bf Testing relativistic electrons in powerful radio-galaxy jets/lobes}
The SZE produced in the jets of isolated powerful radio-galaxies is, similarly to the
case of cluster cavities, completely non-thermal and it is, contrary to the case of
cluster cavities, not contaminated by the surrounding ICM. Colafrancesco (2006a) showed
that high-sensitivity and high-resolution SZE observations (like those achievable with
ALMA) can also provide relevant information on the spectrum and on the pressure and
energetic structure of the jets/lobes of powerful radio-galaxies.

\subsection{SZE and the nature of Dark Matter}

Dark Matter (DM) annihilations in the halo of galaxies and galaxy clusters have relevant
astrophysical implications, even for SZE observations.
Galaxy clusters and dark (dwarf) galaxies are, in fact, gravitationally dominated by Cold
Dark Matter for which the leading candidate is the lightest supersymmetric (SUSY)
particle, plausibly the neutralino $\chi$. Experimental and theoretical considerations
for having a cosmologically relevant neutralino DM lead to bound its mass $M_{\chi}$ in
the range between a few GeV to a few hundreds of GeV (see discussion in Colafrancesco et
al. 2006a). The decays of neutralino annihilation products (fermions, bosons, etc.)
yield, among other particles, energetic electrons and positrons up to energies comparable
to the neutralino mass.
Colafrancesco (2004b) proposed to explore the consequences of the Compton scattering
between the secondary electrons produced from the WIMP annihilation in massive DM halos,
like galaxy clusters and dwarf galaxies, and the CMB photon field, i.e. the DM-induced
SZE which has specific spectral and spatial features. This is an inevitable consequence
of the presence and of the nature of DM in large-scale structures.
\begin{figure}[!ht]
\begin{center}
\hbox{
 \epsfig{file=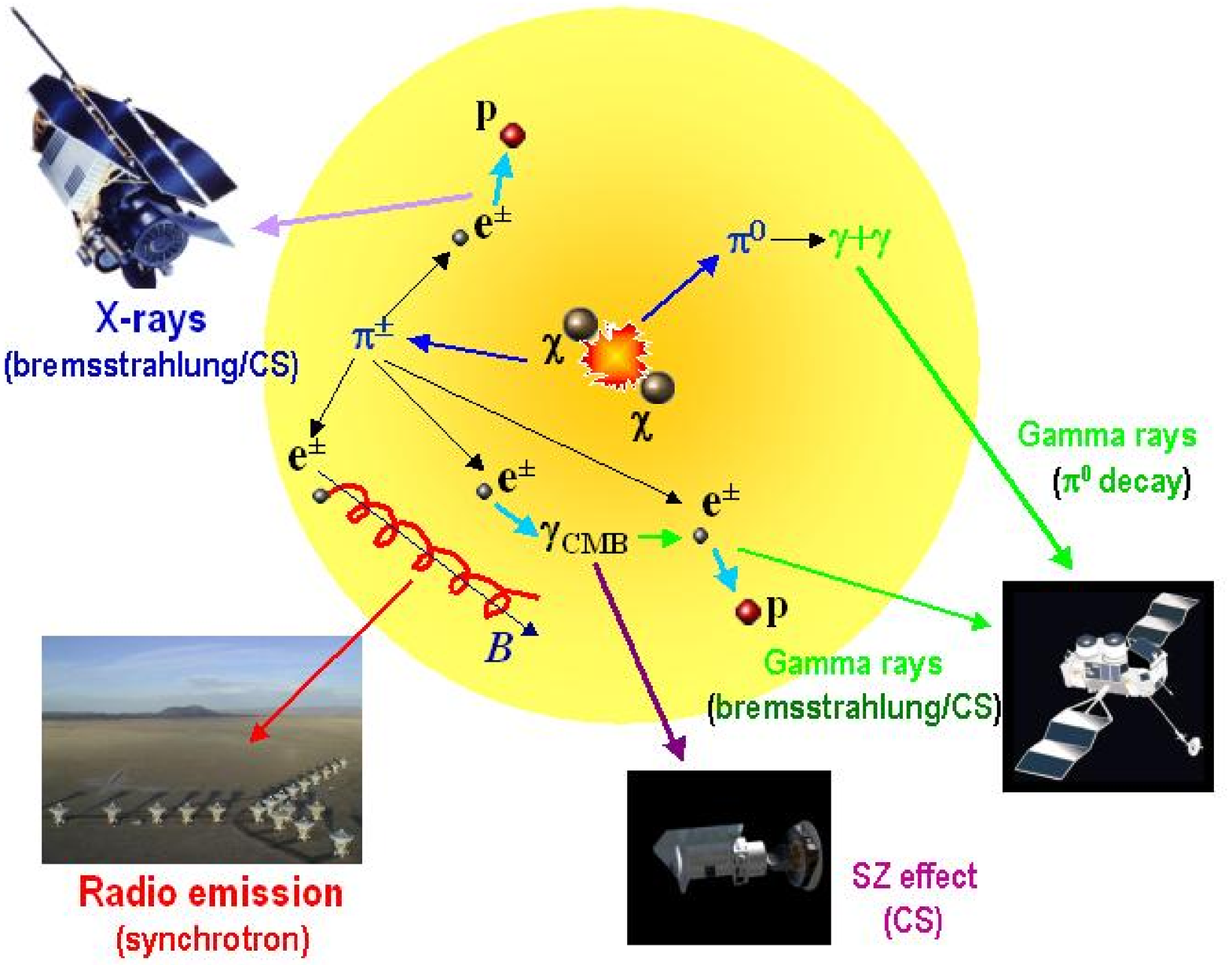,width=8.cm,angle=0.}
 \epsfig{file=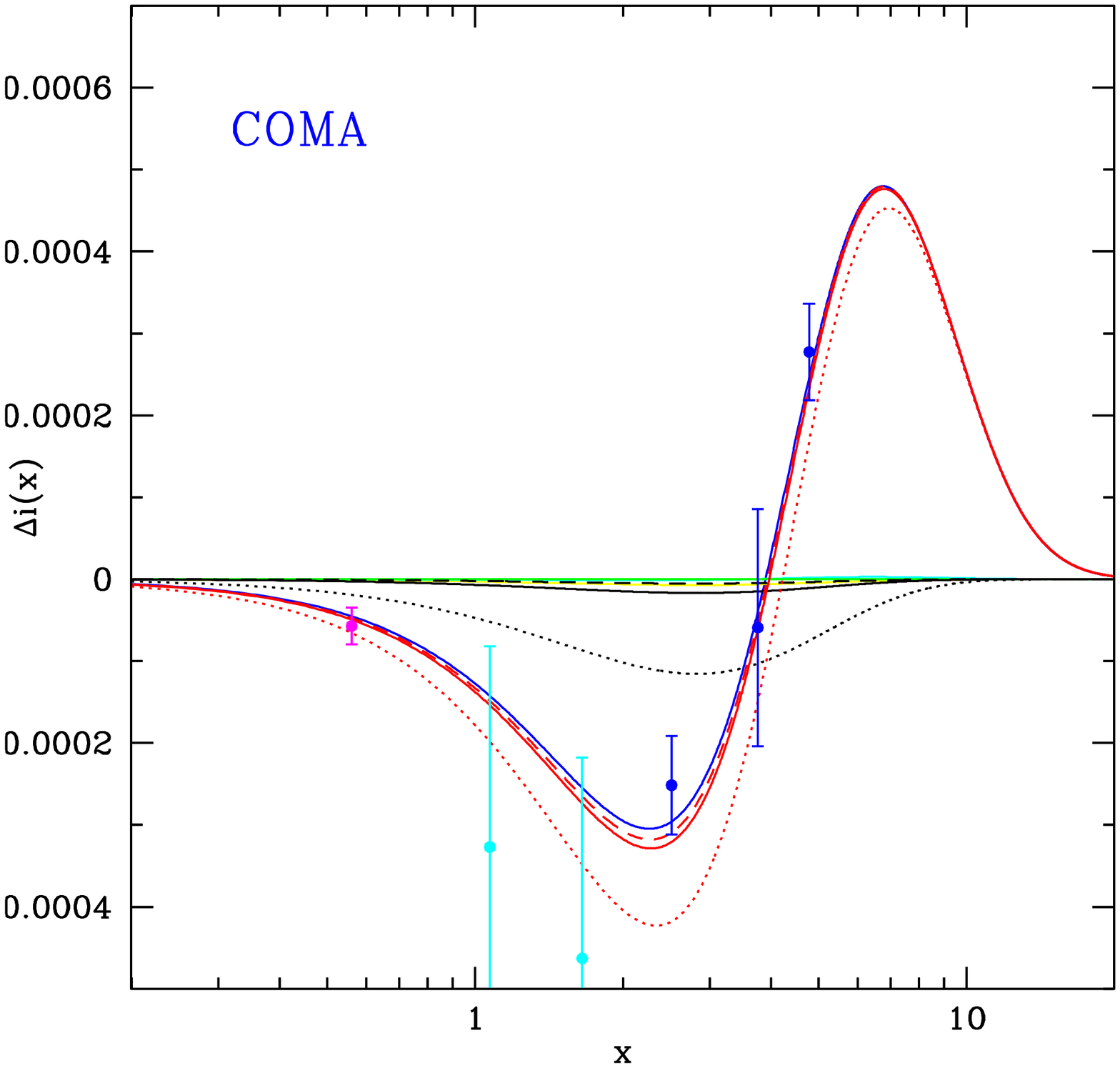,width=6.cm,angle=0.}
  }
  \caption{\footnotesize{Left. A simple model which shows the basic astrophysical mechanisms
  underlying the search for the nature of DM particles ($\chi$) through the emission features occurring in
large-scale structures (e.g., galaxy clusters and galaxies). These mechanisms are, among
others: $\gamma$-ray emission from $\pi^0 \to \gamma+\gamma$, relativistic bremsstrahlung
of secondary $e^{\pm}$ and ICS of CMB photons by secondary $e^{\pm}$; X-ray/UV emission
due to non-thermal bremsstrahlung and ICS of background photons by secondary $e^{\pm}$;
synchrotron emission by secondary $e^{\pm}$ diffusing in the intra-cluster magnetic
field; SZ$_{DM}$ (ICS of CMB photons by secondary $e^{\pm}$) effect.
  Right. The overall SZE in Coma produced by the combination of various electron
  populations: thermal hot gas with $k_{\rm B}T_{\rm e} = 8.2 $ keV and $\tau = 4.9 \cdot 10^{-3}$ (solid blue curve)
  which best fits the available SZ data (DePetris et al. 2002); relativistic electrons which best fit
  the radio-halo spectrum (yellow curve) provide a small additional SZE (Colafrancesco et al. 2003);
  warm gas with $k_{\rm B} T_{\rm e} \approx 0.1$ keV and $n \approx 10^{-3} $ cm$^{-3}$ (cyan curve) provides a small
  SZE due to its low pressure; DM produced secondary electrons with $M_{\chi} = 10$
  (black dotted curve), $20$ GeV (black solid curve) and $30$ GeV (dashed solid curve). A pure-gaugino
  $\chi$ reference model is assumed in the computations.
  The relative overall SZE is shown as the dotted, solid and dashed red curves,
  respectively. A zero peculiar velocity of Coma is assumed consistently with the available limits.
  SZ data are from OVRO (magenta), WMAP (cyan) and MITO (blue).
 }}
  \label{fig.sz_dm}
\end{center}
\end{figure}
The analysis of the DM induced SZE in galaxy clusters provides a probe for the presence
and for the nature of DM in cosmic structures which is complementary to those obtainable
through a multifrequency analysis, from radio to gamma-rays (see Fig.\ref{fig.sz_dm} and
Colafrancesco et al. 2006a). The available SZ observations on the Coma cluster (see
Fig.\ref{fig.sz_dm} and Colafrancesco 2004b) can already set a lower limit to the
neutralino mass of $M_{\chi} \simgt 17 - 20$ GeV ($M_{\chi} \simgt 13$ GeV at 90 \% c.l.
with the adopted value of $\langle \sigma v \rangle_{\rm 0} = 3 \cdot 10^{-27} cm^{-3}
s^{-1}/\Omega_{\chi}h^2$ with $\Omega_{\chi} h^2 = 0.116$), which is consistent with the
limits set by accelerators (e.g., Belanger et al. 2003). The SZ$_{\rm DM}$ signal does
not strongly depend on the assumed DM density profile at intermediate angular distances
from the cluster center and on the DM clumpiness since $y_{\rm DM} = (\sigma T/m_ec^2)
\int \d \ell P_{\rm e,DM}$ is the integral of the total DM-produced secondary electron
pressure, $P_{\rm e, DM}$, along the line of sight.
The presence of a substantial SZ$_{\rm DM}$ effect is likely to dominate the overall SZ
signal at frequencies $x\simgt 3.8-4.5$ providing a negative total SZE (see
Fig.\ref{fig.sz_dm}). It is, however, necessary to stress that in such frequency range
there are other possible contributions to the SZE, like the kinematic effect and the
non-thermal SZE which could provide additional biases (see, e.g., Colafrancesco et al.
2003).
Nonetheless, the peculiar spectral shape of the $SZ_{\rm DM}$ effect is quite different
from that of the kinematic SZE and of the thermal SZE and this allows to disentangle it
from the overall SZ signal.
An appropriate multifrequency analysis of the overall SZE based on observations performed
on a wide spectral range (from the radio to the sub-mm region) is required, in principle,
to separate the various SZ contributions and to provide an estimate of the DM induced
SZE.
In fact, simultaneous SZ observations at $\sim 150$ GHz (where the SZ$_{\rm DM}$ deepens
the minimum with respect to the dominant thermal SZE), at $\sim 220$ GHz (where the
SZ$_{\rm DM}$ dominates the overall SZE and produces a negative signal instead of the
expected $\approx$ null signal) and at $\simgt 250$ GHz (where the still negative
SZ$_{\rm DM}$ decreases the overall SZE with respect to the dominant thermal SZE) coupled
with X-ray observations which determine the gas distribution within the cluster (and
hence the associated dominant thermal SZE) and lensing data (which determine the shape of
the DM-dominated potential wells) can separate the SZ$_{\rm DM}$ from the overall SZ
signal, and consequently, set constraints on the neutralino mass. Observations of the
radio-halo emission in the cluster can provide an estimate of the cosmic-ray electron
population and consequently an estimate of the associated non-thermal SZE (which is
usually quite small and with a different spectral shape at high frequencies, see e.g.,
Colafrancesco et al. 2003).
A particularly good case in which the SZ$_{DM}$ could be revealed is that of the cluster
1E0657-556 (see Fig.\ref{fig.clusters}) where the SZ$_{DM}$ signal is peaked on the DM
clumps, and is well separated (by several arcmin) from the thermal SZE concentrated on
the X-ray emitting IC gas location and from the non-thermal SZE associated with the
shock. We stress here that the SZ$_{DM}$ signal is the only one remaining at frequencies
$x=x_{o,th}$ at the DM clump locations (see Colafrancesco et al. 2006c).
Because the amplitude of the SZ$_{\rm DM}$ effect increases with decreasing values of
$M_{\chi}$, the high-sensitivity SZ experiments have, hence, the possibility to set
reliable constraints to the nature, amount and spatial distribution of DM in galaxy
clusters.

While the SZ$_{\rm DM}$ effect in galaxy clusters could be contaminated by other possible
sources of SZE (thermal from both hot and warm gas, non-thermal from cosmic rays,
relativistic electrons injected by radio-galaxies and AGNs, kinematic due to the cluster
peculiar motion), the Comptonization of the CMB photons by secondary electrons produced
by DM annihilation in a "pure" DM halo (i.e., a dark galaxy) produce a SZ$_{\rm DM}$
which is un-contaminated and therefore carries secure information on the DM nature.
Colafrancesco (2004b, 2005b) proposed that this is the case to be expected in a dwarf
galaxy like Draco where no material other than DM and stars are present. In this respect,
the possible detection of the SZ$_{\rm DM}$ from nearby dwarf galaxies could either
provide extremely clear and strong constraints on the DM nature or even detect their
indirect signals.
Detailed analysis on the SZE expected from Draco (Culverhouse, Evans \& Colafrancesco
2006, Colafrancesco, Profumo \& Ullio 2006b) revealed that the SZ$_{\rm DM}$ signal is
however quite tiny, and typically requires long duration observations with the future
sensitive SZ experiments like ALMA.
However, future SZ experiment with sub-$\mu$K sensitivity (and even moderate angular
resolution) could be able to detect direct signals from DM annihilation in the most
favourable supersymmetric scenarios.

\subsection{SZE and magnetic fields}

There are also non-standard aspects of the widely-used thermal SZE that need to be
carefully and specifically addressed. In particular, the presence of an intra-cluster
magnetic field can produce relevant changes to the thermal SZE in galaxy clusters (see,
e.g., Colafrancesco \& Giordano 2006a for details).
We know, in fact, that magnetic fields exist in clusters of galaxies for several reasons:
first, in many galaxy clusters we observe the synchrotron radio-halo emission produced by
relativistic electrons spiraling along magnetic field lines; second, the Faraday rotation
of linearly polarized radio emission traversing the ICM proves directly and independently
the existence of intracluster magnetic fields (see, e.g., Carilli \& Taylor 2002, Govoni
\& Feretti 2004 for reviews). Other estimates of the magnetic field strength on the
cluster wide scale come from the combination of synchrotron radio and inverse Compton
detections in the hard X-rays (e.g., Colafrancesco, Marchegiani \& Perola 2005), from the
study of cold fronts and from numerical simulations (see, e.g., Govoni \& Feretti 2004).
This evidence provides indication on the wide-scale B-field which is at the level of a
few tens up to several $\mu$G (and in some cases up to $\sim 10$ $\mu$G, as in Coma) with
the larger values being attained by the most massive systems.
\begin{figure}[!ht]
\begin{center}
\hbox{
 \epsfig{file=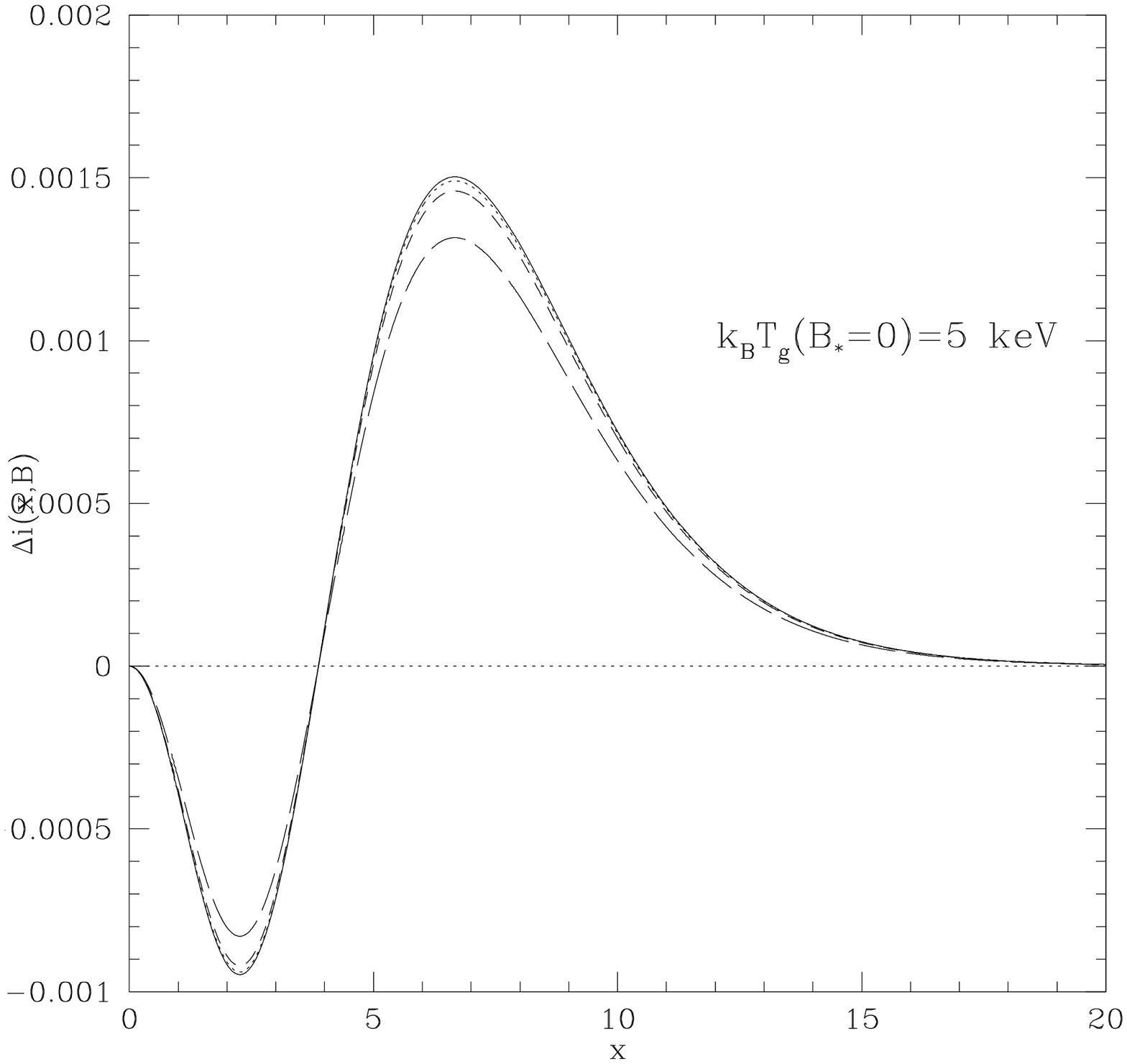,width=6.cm,angle=0.}
 \epsfig{file=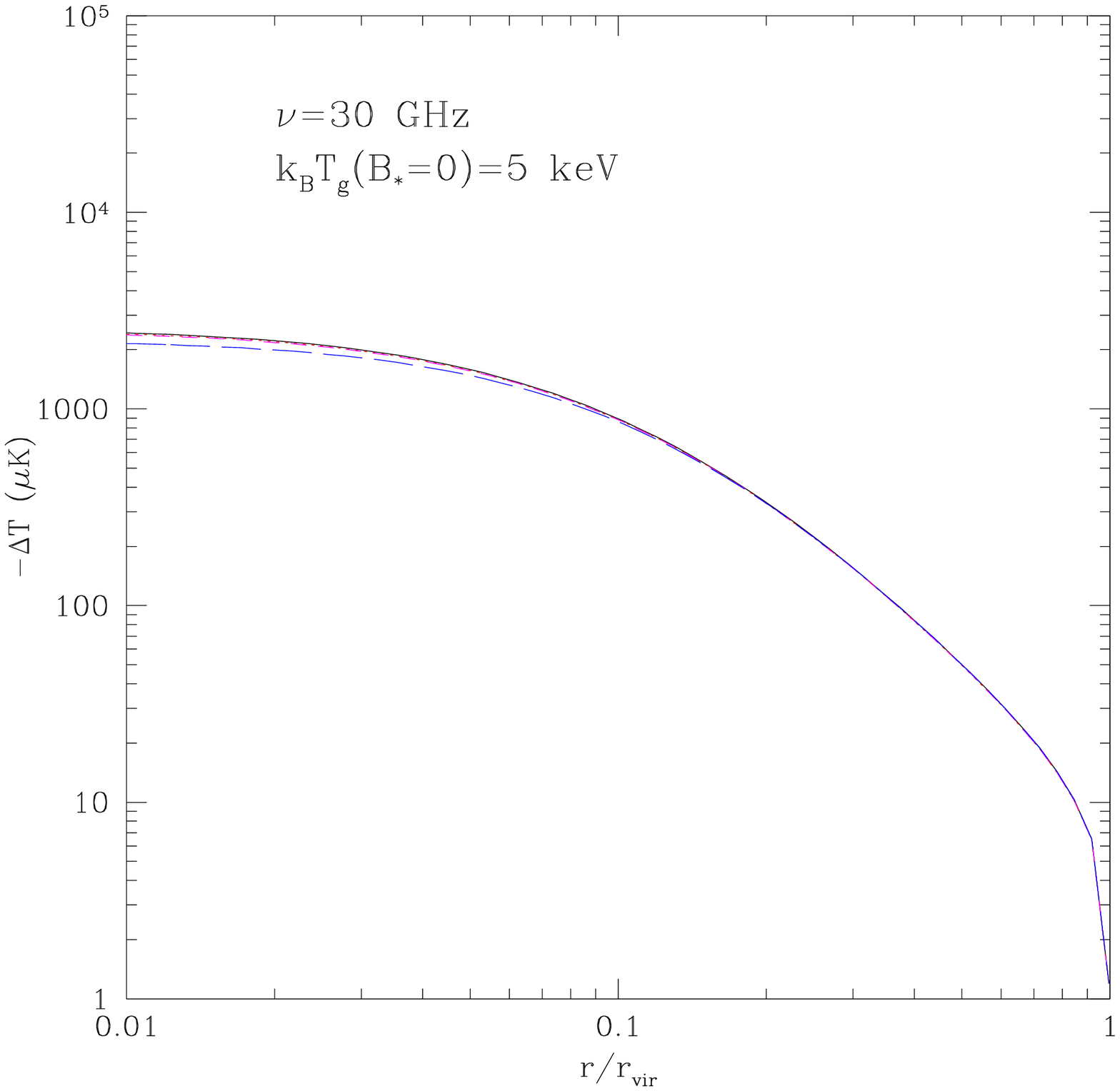,width=6.cm,angle=0.}
  }
  \caption{\footnotesize{The modifications of the spectral (left) and spatial
  distribution (right) of the thermal SZE for a $k_BT_e = 5$ keV cluster
  induced by a magnetic field $B(r) \propto B_* \rho_g^{0.9}$ for values $B_*=$ 0 (solid),
  0.5 (dotted), 1 (dot-dashes) and 3 (dashes) $\mu$G. Figure from Colafrancesco \&
  Giordano (2006a).
 }}
  \label{fig.sz_b}
\end{center}
\end{figure}
The presence of a wide-scale magnetic field implies modification on the thermal and
density structure of the IC gas by acting on both the magnetic virial theorem (MVT, from
which the overall IC gas temperature is derived, see Colafrancesco \& Giordano 2006b)
  \be
 2U + U_B + W = 0 \, ,
 \label{eq.mvt}
 \ee
(here $U$ is the kinetic energy, $W$ is the potential energy and $U_B$ is the magnetic
energy of the system), and the hydrostatic equilibrium condition (HE, from which the
radial profile of the IC gas density is derived, see Colafrancesco \& Giordano 2006c)
 \be
\frac{dP_g}{dr}+\frac{dP_B}{dr}=-\frac{GM(\leq r)}{r^2}\rho_g,
 \label{eq.he}
 \ee
where $\rho_g \equiv \rho_g(r,B)$ is the density of the thermal IC gas and $P_g(r,B)$ is
its pressure, with both quantities depending in general on the magnetic field $B$.
As a consequence, the cluster thermal SZE (see eqs. 1-4 and 7-11) is certainly influenced
by the presence of an intracluster magnetic field, as shown in Fig.\ref{fig.sz_b}. For
idealized isothermal clusters, the magnetic SZE is reduced w.r.t. the unmagnetized case
by two separate effects: i) the decrease in the cluster temperature as derived by the MVT
(see eq.\ref{eq.mvt}) for increasing values of $B$; and ii) the decrease of the central
IC gas density profile according to the HE condition (see eq.\ref{eq.he}) for increasing
values of $B$.
The variations of the thermal SZE for magnetized clusters is larger for lower-mass
systems because in these structures the pressure provided by the magnetic field is of
larger relative importance w.r.t. to the thermal pressure of the IC gas settling in the
potential wells provided by Dark Matter.
The increasing observational evidence, the refinements of numerical simulations and the
theoretical expectations for the presence of magnetic fields in galaxy clusters render a
revision of the standard description of the thermal SZE necessary to describe both single
SZ observations and SZ scaling-law analyses in a self-consistent astrophysical and
cosmological framework (see discussion in Colafrancesco \& Giordano, 2006a).

\section{Epilogue}
 \label{sec.outline}

The most recent achievements in multi-frequency (radio thru X-rays) precision
observations of galaxies and galaxy clusters provided a wealth of detailed and, in some
cases, unexpected physical information on the structure, physical state and composition
of the atmospheres of these cosmic structures.
Radio-halo non-thermal emission, hard--X-ray quasi-thermal excess emission, cluster
cavities and radio bubbles, absence of strong radiative cooling in cool-cores, AGN's
feedback, Dark Matter distribution and dynamics, appear to be relevant ingredients for a
detailed description of cluster structure and evolution: these processes put the standard
description of the cluster atmosphere (as a single, thermal IC gas) to the ropes.
Therefore, the theoretical modelling of galaxy cluster atmospheres evolved in the last
decade up to a level of high-detail, complex physical description that requires to take
consistently into account several electronic components (of thermal, non-thermal and
relativistic nature), the effects of magnetic field, the feedback (heating, particle
injection, magnetic field compression, etc.) produced by AGNs, radio-galaxies and
blazars, the evolution and the interaction of relativistic plasma bubbles, the complex
interplay of cooling and (non-thermal) heating processes in cool cores, the effects of a
possible Dark Matter annihilation, the non-trivial physical conditions of the IC gas at
cluster boundaries and its transition into the cosmic-web IGM distribution on larger
scales.\\
As a consequence, the standard lore of the SZE is no longer viable in cosmic structures
on large scales, like galaxy clusters. The simple SZ physics is no longer representative
of the actual observational status; in this sense, it cannot provide neither reliable
cluster physics nor an adequate cosmological use.\\
It is therefore inevitable to go beyond the standard lore of the SZE. This provides,
nonetheless, a way to use SZE as a single technique to efficiently study the leptonic
structure of clusters/galaxy atmospheres and thus obtain information on the density,
entropy, pressure and energetics of the electrons, the presence and the spectra of
different electron populations, their (possible) equilibrium conditions.\\
The SZ signals expected from the non-thermal plasma, from DM annihilation and from
cluster cavities and radio bubbles in addition to the thermal SZE and its modifications
due to the inclusion of the intra-cluster magnetic field physics are usually in the range
from a few $\mu$K to a few tens of $\mu$K. This is the reason why they have not yet
clearly discovered in the past and ongoing SZ experiments.
These additional SZ components are however present in the overall SZ signal observable
from many galaxy clusters, and therefore they enrich the available physical information
contained in the SZE data. Such signals could be disentangled from the thermal SZE by the
future experiments with $\mathcal{O} (\mu K)$ sensitivity and sub-arcmin resolution.
Nonetheless, they constitute a bias of complex spatial and spectral nature for those
experiments with $\simgt 10 \mu K$ sensitivity (like WMAP and PLANCK). In each case,
their theoretical study and simulation analysis is mandatory for any astrophysical and
cosmological use to be reliably carried on.\\
Such a program requires a definite technological effort which is directed towards high
sensitivity (sub-- or $\sim \mu$K level) and high spatial resolution ($\sim$
arcsec--arcmin level) together with a wide-band continuum spectral coverage obtainable
from space experiments (see Masi et al. 2006). This goal is definitely at the frontier of
the present technology and is therefore a challenge for the future SZE experiments but it
will be, nonetheless, able to open the door to the exploration of fundamental
(astro-particle) issues like the nature of Dark Matter, the origin and the distribution
of cosmic rays and magnetic fields in large-scale structures and other relevant questions
which are on the discussion table of modern astrophysics and cosmology.



\end{document}